\documentclass[aps,preprint,amsfonts,amsmath]{revtex4}

\usepackage{graphicx}  % Graphics capability
\begin{document}
\author{Martin J.~Greenall}
\affiliation{Institut Charles Sadron, 23, rue du Loess, 67034 Strasbourg, France}
\author{Peter Schuetz}
\affiliation{Unilever R\&D Colworth, Colworth Park, Sharnbrook, MK44 1LQ, UK}
\email{peter.schuetz@unilever.com}
\author{Steve Furzeland}
\affiliation{Unilever R\&D Colworth, Colworth Park, Sharnbrook, MK44 1LQ, UK}
\author{Derek Atkins}
\affiliation{Unilever R\&D Colworth, Colworth Park, Sharnbrook, MK44 1LQ, UK}
\author{D.~Martin A.~Buzza}
\affiliation{Department of Physics, The University of Hull, Cottingham Road, Hull HU6 7RX, UK}
\author{Michael F. Butler}
\affiliation{Unilever R\&D Colworth, Colworth Park, Sharnbrook, MK44 1LQ, UK}
\author{Tom C. B. McLeish}
\affiliation{Department of Physics, Durham University, South Road, Durham DH1 3LE, UK}
\title{Controlling the self-assembly of binary copolymer mixtures in
  solution through molecular architecture}
\begin{abstract}
We present a combined experimental and theoretical study on the role
of copolymer architecture in the self-assembly of binary PEO-PCL
mixtures in water-THF, and show that altering the chain geometry and
composition of the
copolymers can control the form of the self-assembled
structures and lead to the formation of novel aggregates. First,
using transmission electron microscopy and turbidity measurements, we
study a mixture of sphere-forming and lamella-forming PEO-PCL copolymers,
and show that increasing the molecular weight of the lamella-former at a constant ratio of its
hydrophilic and hydrophobic components leads to the formation of
highly-curved structures even at low sphere-former
concentrations. This result is explained using a simple argument based
on the effective volumes of the two sections of the diblock and is
reproduced in a coarse-grained
mean-field model: self-consistent field theory (SCFT). Using further
SCFT calculations, we study the distribution of the two
copolymer species within the individual aggregates and discuss how
this affects the self-assembled structures. We also
investigate a binary mixture of lamella-formers of different
molecular weights, and find that this system forms vesicles with
a wall thickness intermediate to those of the vesicles formed by the two
copolymers individually. This result
is also reproduced using SCFT. Finally, a mixture of sphere-former and
a copolymer with a large hydrophobic block is shown to form a range of structures, including novel
elongated vesicles.
\end{abstract}
\maketitle
\section{Introduction}
Amphiphiles such as block copolymers and lipids can self-assemble into many
different structures when dissolved in solution \cite{jain_bates,jain_bates_macro}. The case of block copolymers
has proved especially interesting to researchers in recent years, for
a variety of reasons \cite{smart}. First, the study of block copolymers is a promising route to a
fundamental model of the self-assembly of amphiphiles in solution, since the
theoretical understanding of the constituent polymer molecules is on a firm
footing. Well-established methods such as self-consistent field theory
(SCFT) \cite{edwards,schmid_scf_rev,matsen_book}
have provided considerable insight into the self-assembly of polymers,
especially in melts \cite{maniadis,drolet_fredrickson}, whilst using simple models of the individual polymer
molecules. Second, vesicles formed from block copolymers show more promise
as vehicles for drug delivery \cite{kim} than similar structures formed from lipids, as
the thickness and low solubility of their membranes means that they can be
longer-lived and less permeable \cite{discher,discher_eisenberg}.

For solutions of a single type of diblock copolymer, it is often relatively
straightforward to understand why a given type of aggregate forms in a given
system. The main factor that determines the shape of the structures is the
architecture of the copolymer; that is, the size of its hydrophilic and
hydrophobic blocks \cite{kinning_winey_thomas} (although other
factors, such as the overall size of the copolymer, may also play a role \cite{kaya}). If the hydrophilic component is large compared to the
hydrophobic component, then curved aggregates such as spherical or cylindrical
micelles form. Conversely, if the hydrophobic component is large, lamellar
structures such as vesicles are observed
\cite{kinning_winey_thomas}. Recently, we demonstrated this behavior in a
study of PCL-$b$-PEO block-copolymers with various volume fractions of
the hydrophobic block (PCL) \cite{adams}. Here, large volume fractions $f_\text{EO}$ of the 
hydrophilic block (PEO) resulted in micelles ($f_\text{EO}>0.3$), lower $f_\text{EO}$
favored wormlike micelles ($0.25<f_\text{EO}<0.3$) and still lower
fractions ($f_\text{EO}<0.25$) led to the formation of vesicles.

We can gain increased control over the self-assembly by mixing two types of
amphiphile that individually form aggregates of different curvatures. Such
mixtures are well known in cell biology, where different lipids can be sorted
by segregation to regions of high and low curvature \cite{sorre,zidovska}, and have also
been studied in the context of lipid-detergent systems
\cite{vinson,oberdisse}. More recently, they have been
investigated in block copolymer solutions. For example, Jain and Bates
\cite{jain_bates_macro} have
studied mixtures of polyethylene oxide-polybutadiene (PEO-PB), and have found
that blending ratio can be used to control self-assembly, and furthermore that
novel structures such as undulating cylinders form. They also found that
different aggregates form depending on whether the two polymer species are
mixed before or after their individual self-assembly \cite{jain_bates_macro}.

In a recent study of a mixture of sphere-forming and lamella-forming
polycaprolactone-co-polyethylene oxide in water-THF mixed solvents \cite{schuetz}, we have built on this work by
controlling the quantities of water and THF to mix sphere- and
lamella-forming copolymers not only
before and after but also during their individual self-assembly. Those copolymers mixed before
self-assembly (pre-mixed) formed a sequence of aggregates of increasing
curvature as the amount of sphere-former was increased, forming vesicles, then
a mixtures of vesicles, rings and worms, and finally spherical
micelles. This series of shape transitions has also been observed in
lipid-detergent mixtures \cite{vinson,oberdisse}, and has been studied
theoretically using self-consistent field theory \cite{schuetz,li,gg} and
models of chain packing \cite{fattal} and membrane curvature \cite{andelman}. 
When
mixed after self-assembly (post-mixed), the two species remained
locally in the equilibrium states of the pure components, and a mixture of vesicles and spherical micelles was observed.
The structures found when the two species were allowed partially to
self-assemble before mixing (intermediate mixing) were more unusual, and
included metastable paddle- and horseshoe-shaped aggregates. Using self-consistent field theory, we
reproduced the transitions between morphologies observed in the pre-mixed
system and also details of the aggregates such as the bulbous ends of
the rods \cite{schuetz}. We also gained insight into the complex
metastable structures seen at intermediate
mixing, by showing in SCFT calculations that the segregation of the two types
of copolymer can stabilize regions of different curvature within a single
aggregate.

In the current paper, we extend this study by varying the architectures of the
copolymer species. We consider three specific cases: varying the
length of the lamella-former in a blend of sphere- and
lamella-formers, blending two lamella-formers of different lengths,
and blending sphere-former with a polymer that has such a large
hydrophobic block that it precipitates in solution if not mixed with more
hydrophilic molecules. In all cases, we observe the quantitative and qualitative changes in the
self-assembly as the copolymer architectures are changed. As in our
previous work \cite{schuetz}, we
perform self-consistent field theory calculations in tandem with our
experiments and
discuss how the distribution of the two copolymer species within the
self-assembled structures leads to the formation of the structures seen in the
experiments.

The article is organised as follows. In the following section, we give
details of our experimental and theoretical methods. The Results
section is divided into three subsections, one for each of the
mixtures introduced above. We then present our conclusions.

\section{Methods}\label{methods}
\subsection{Materials}
The PEO-PCL block copolymers were purchased from Advanced Polymer Materials
Inc., Montreal and used as received.  GPC analysis was also provided by Advanced
Polymer Materials Inc. and was referenced against PEO standards. Degrees of
polymerization for the PCL block were calculated by $^1\text{H}$ NMR in $\text{CDCl}_3$ by comparison to the PEO block (the degrees of polymerization for the
monomethoxypoly(ethylene oxides) used in these polymerizations are known).
The molecular weight and molecular weight distributions are given in
Table \ref{tab:scattering}.
All other reagents with the exception of NMR solvents were purchased from Sigma
Aldrich Company Ltd., Gillingham. Standard solvents were of
spectrophotometric grade and inhibitor free.  Deuterated NMR solvents were
purchased from Euriso-top S.A., Saint-Aubin. All solvents were filtered before
use through Pall Acrodisc PSF GHP $200\,\text{nm}$ filters.  For all
experiments, distilled and de-ionized Millipore water (resistivity =
$18.2\,\text{M}\Omega.\text{cm}$) was additionally filtered through Sartorius
Ministart $200\,\text{nm}$ filters directly before use.

\begin{table}
\begin{center}
\begin{tabular}{|c|c|c|c|c|c|c|}
\hline
Commercial sample code & Sample formula & $M_w\,^\text{a}$ & $M_w/M_n$ & $f_\text{EO}\,^\text{b}$ & Morphology $^c$ & $R_{h}\,(\text{nm})\,^\text{d}$ \tabularnewline \hline
PCL$_\text{10k}$PEO$_\text{2k}$ & PEO$_\text{45}-b-$PCL$_\text{101}$ & 17300 & 1.36 & 0.15 & V & 220 \tabularnewline \hline
PCL$_\text{5k}$PEO$_\text{1k}$ & PEO$_\text{23}-b-$PCL$_\text{47}$ & 7100 & 1.15 & 0.17 & V & 170 \tabularnewline \hline
PCL$_\text{5k}$PEO$_\text{2k}$ & PEO$_\text{45}-b-$PCL$_\text{43}$ &
7800 & 1.16 & 0.3 & C,S & 30 \tabularnewline \hline
PCL$_\text{5k}$PEO$_\text{550}$ & PEO$_\text{12}-b-$PCL$_\text{56}$ &
6380 & 1.15 & 0.08 & P & n/a \tabularnewline \hline
\end{tabular}
\end{center}
\caption{\label{tab:scattering} {\bf Polymers used in this study, with their properties from light scattering.}  Notes: (a) total molecular weight from GPC (PEO standards) (b) volume fraction of the EO block calculated from the melt densities of the two blocks (c) morphology determined by light scattering and cryo-TEM (S=spherical micelles, C = worm-like micelle, V = vesicle, P = precipitate; in the cases of mixed morphologies the majority component is written first), (d) hydrodynamic radius from dynamic light scattering (DLS) after dialysis.}
\end{table}
\subsection{Preparation of Solutions}  Aqueous dispersions of block-copolymer
aggregates were prepared by dissolving the polymer in THF to a concentration
of $10\,\text{mg ml}^{-1}$.  These solutions were then mixed in the volume ratio noted for
the experiments. All the mixing ratios are thus ratios of the masses
of the respective polymers as opposed to molar ratios. Due to the close
molecular weights (Table \ref{tab:scattering}) of the two copolymers PCL$_\text{5k}$PEO$_\text{1k}$ and PCL$_\text{5k}$PEO$_\text{2k}$
these ratios are not very different in this case, while for the other combinations the conversion is easily calculated. Millipore water was added either
manually or by an Eppendorf EDOS 5222 Electronic Dispensing System. 125
aliquots of $20\,\mu\text{l}$ of water were added at one-minute intervals.
\subsection{Turbidity Measurements}
We performed
turbidity measurements during the preparation of the samples 
using an adapted Perkin Elmer UV/Vis
Lambda 40 Spectrometer.  A wavelength of $600\,\text{nm}$ was used with a slit width of
$2\,\text{nm}$.  Stirring was performed using a standard magnetic stirrer/hotplate
placed under the spectrometer.  The polymer was dissolved in THF
($1\,\text{mL}$, $10\,\text{mg mL}^{-1}$) and a zero reading was taken (transmittance, $T = 100\%$).
Millipore water was then added either in $10\,\mu\text{l}$ aliquots every $30\,\text{s}$ using an
Eppendorf EDOS 5222 Electronic Dispensing System and a turbidity
reading was taken
after each addition. 
\subsection{Cryo-TEM}
Samples for thin-film cryo-TEM were loaded onto
plasma-treated (30 seconds) holey-carbon grids and prepared using a GATAN
cryo-plunge into liquid ethane and then transferred using a GATAN 626
cryo-transfer system.  Samples were examined using a JEOL 2100 TEM operating
at $200\,\text{kV}$.  Images were obtained using a Bioscan or a GATAN Ultrascan 4k camera
and analyzed by GATAN Digital Micrograph version 1.71.38. During our previous investigations \cite{adams} we observed that
imaging of the self-assembled structures (especially vesicles) is greatly
improved in samples containing ca.\ $30\%$ THF compared to samples
in pure aqueous solution. Similar structures were observed in
both solvent conditions, which indicates that at THF fractions of
$30\%$ and below, the mobility of the block-copolymers is too
restricted to allow for further growth of the aggregates. However, we
found that, over a timescale of a few months, internal rearrangements
occurred that evened out local variations in surface curvature
transforming the more complex metastable aggregates into vesicles or
nested onion-like structures. In order to focus on the initial metastable structures that
are formed (prior to any slow internal rearrangement processes) and to
obtain as high quality imaging as possible, we prepared the solutions
for cryo-TEM in aqueous solutions containing $28\%$ THF and imaged
these samples within a maximum time of two weeks.

\subsection{Self-consistent field theory} 
To further our understanding of
the role of polymer architecture on
self-assembly, we performed
self-consistent field theory (SCFT) calculations \cite{edwards} on a model system of two
species of AB diblock copolymers (lamella- and sphere-forming respectively)
blended with A homopolymer `solvent'. SCFT is a coarse-grained
mean-field theory
in which the individual polymer molecules are modeled by random walks
and 
composition fluctuations are neglected. For sufficiently long polymers
\cite{muller_book}, these approximations prove extremely
effective and the predictions of the theory are very accurate for a
wide range of systems \cite{matsen_book}. 
We use a simple implementation
of the theory where the interactions of the polymers are
included by imposing incompressibility and introducing a contact potential
between the A and B monomers \cite{matsen_book}. The fine details of
the polymer molecules are not taken into account, so that monomers of all
species are taken to have the same length $a$ and volume
$1/\rho_0$. 

From a technical point of view, a self-consistent field theory
calculation consists of solving a series of differential (diffusion) equations to
calculate the density profiles of the various polymer species.
An initial guess for the profiles is made, which has the approximate
form of the structure that we wish to study. The density profiles are then recalculated until a set of equations reflecting the
physical properties of the system (such as its incompressibility)
\cite{matsen_book,matsen_jcp} is satisfied. The SCFT differential equations are solved using a finite-difference
method \cite{num_rec} with a spatial step size of
$0.04\,aN^{1/2}$, where $N$ is the number of monomers in the
sphere-forming species, and reflecting boundary conditions are imposed at the origin and
edges of the box. Full technical details of our calculation can be found in a
recent publication \cite{gg}.

In the current paper, we use SCFT calculations in two slightly
different ways. First, we perform effectively one-dimensional calculations
on spherical micelles and infinite cylinders and bilayers to calculate
simple phase diagrams as a function of sphere-former volume
fraction and reproduce the basic phenomenology observed in the experiments. These calculations proceed as follows
\cite{gbm_macro,gbm_jcp}. To begin, we
calculate the free-energy
density of a box containing a single spherical, cylindrical or planar
aggregate surrounded by solvent. The shape of this box is set by the symmetry of the aggregate; for example, a spherical micelle is formed at the center of a spherical box. The calculation is therefore effectively one-dimensional.
The volume $V$ of this simulation box is then varied, keeping the volume fraction of copolymer
constant, until the box size with the minimum free-energy density is found.
Provided the system is dilute, so that each aggregate is surrounded by a
large volume of solvent and the aggregates do not interact with each other, this provides a simple model of a larger system (of fixed
volume and fixed copolymer volume fraction) containing many
aggregates. The reason for this is that such a system minimizes its
free energy by varying the number of aggregates and hence the volume
(`box size') occupied by each. Although computationally inexpensive,
this approach yields accurate information on micelle shape
transitions and its results agree well with experiment \cite{gbm_jcp}.

We also carry out more detailed calculations on the rod and ring
structures seen in the experiment. We have two aims
here. First, we wish to show that these more complex structures can
be reproduced in detail in our calculations. Second, we will study
the distribution of the sphere-forming and lamella-forming copolymers
within the aggregates. Since both rods and rings have cylindrical
symmetry, we will perform our (effectively two-dimensional)
calculations in a cylindrical box. We note that it is
not possible to include information on the distribution of the sphere-
and lamella-formers within the micelles in the initial guess for the SCFT
calculations. Any segregation of the two species will therefore arise
naturally from the theory and need not be artificially introduced \cite{gg}.

Although this model is relatively simple, we found in our
previous work on the self-assembly of binary PEO-PCL mixtures
\cite{schuetz} that it contains enough detail to yield information on
the structures formed in such systems and on the distribution of the
two polymer species within these aggregates. In this earlier paper, we focused on a mixture of
lamella-forming PCL$_\text{5k}$PEO$_\text{1k}$ and sphere-forming
PCL$_\text{5k}$PEO$_\text{2k}$. We modeled the lamella-former by a
symmetric AB diblock with equal numbers of monomers $N_\text{A}$ and
$N_\text{B}$ in its hydrophilic (A) and hydrophobic (B) sections. For
simplicity, the homopolymer was taken to have the same length as the
lamella-former. In line with the experiments, the sphere-former
contained the same number of hydrophobic monomers as the
lamella-former but a larger number of hydrophilic monomers, so that
$N_\text{A}=3N_\text{B}$. The $\chi$ parameter setting the strength of
the interaction between the A and B species was set to $\chi=30/N$,
where $N$ is the total number of monomers in the sphere-forming
species. Our aim here was not to match the
experimental polymer parameters exactly, but to reproduce the basic
phenomenology of the system (sphere- and lamella-forming species,
matched hydrophobic blocks) as simply as possible. We take a similar
approach in the current paper. To study the effect of lamella-former
length on the blend of sphere- and lamella-former, and to observe the
result of blending two different lamella-formers, we introduce the larger
lamella-forming copolymer PCL$_\text{10k}$PEO$_\text{2k}$. We model
this new molecule by an SCFT polymer with $3N/4$ monomers,
while keeping $N_\text{A}=N_\text{B}$. The number of monomers $3N/4$
is chosen since increasing the size of the symmetric copolymer too
much will lead to its forming micelles rather than bilayers
\cite{kinning_winey_thomas, kaya}. For the sake of clarity, we
summarize the SCFT polymer parameters introduced above in Table \ref{tab:SCFTpoly}.

\begin{table}
\begin{center}
\begin{tabular}{|c|c|c|c|}
\hline
Polymer & Sample code & SCFT monomers & SCFT $N_\text{A}/N_\text{B}$ \tabularnewline \hline
lamella-former (long) & PCL$_\text{10k}$PEO$_\text{2k}$ & 3N/4 & 1 \tabularnewline \hline
lamella-former (short) & PCL$_\text{5k}$PEO$_\text{1k}$ & N/2 & 1 \tabularnewline \hline
sphere-former & PCL$_\text{5k}$PEO$_\text{2k}$ & N & 3 \tabularnewline \hline
solvent & n/a & N/2 & n/a \tabularnewline \hline
\end{tabular}
\end{center}
\caption{\label{tab:SCFTpoly} {\bf Parameters of the polymers used in
    SCFT calculations.} For each SCFT polymer, we list the
  corresponding polymer in the experiments, the number of (SCFT)
  monomers and the volume ratio of the hydrophilic and hydrophobic blocks.}
\end{table}
\section{Results}\label{results}
\subsection{Blends of lamella- and sphere-formers}
\begin{figure}
\includegraphics[width=\linewidth]{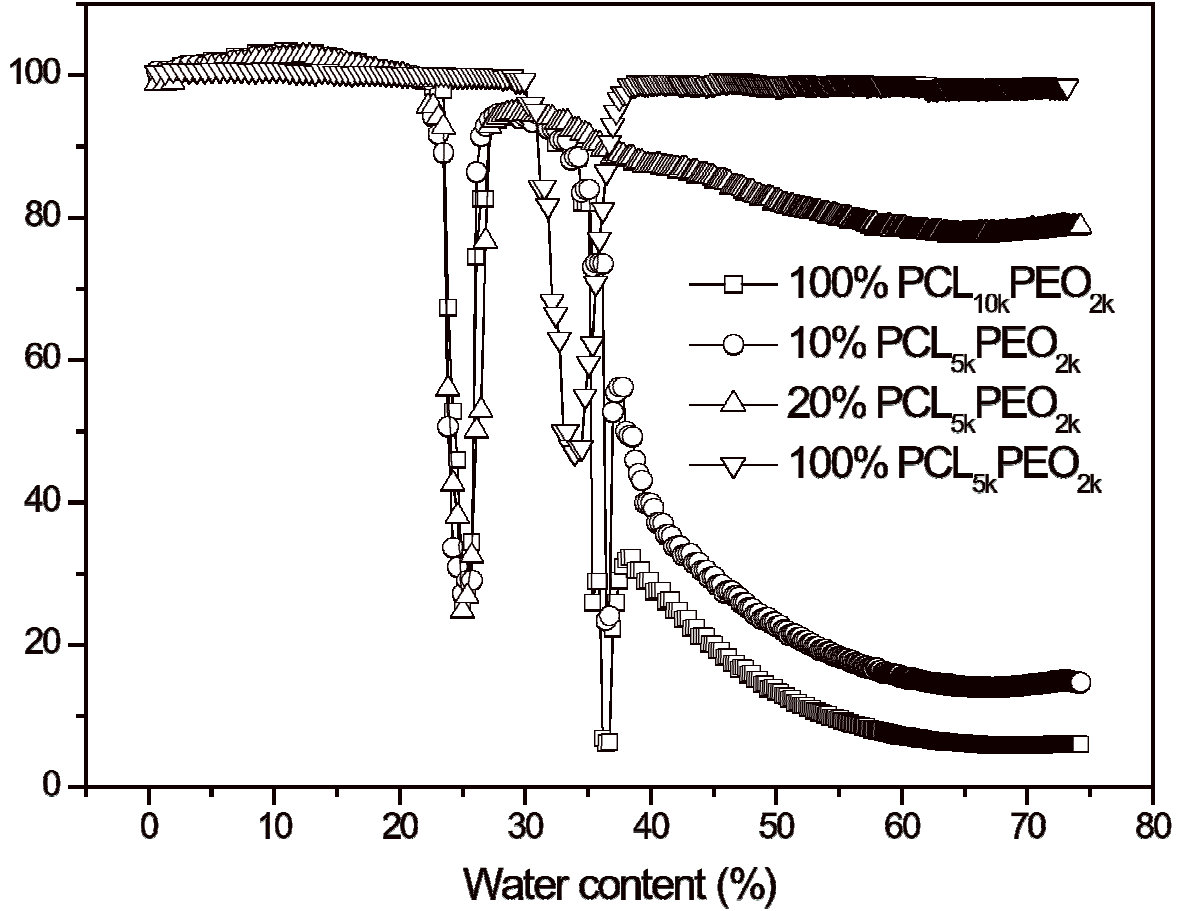}
\caption{\label{turbidity1_fig} Turbidity traces for the self-assembly of block copolymer mixtures of PCL$_\text{10k}$PEO$_\text{2k}$ and PCL$_\text{5k}$PEO$_\text{2k}$. The optical transmission (in \%) at $600\,\text{nm}$ is plotted against the water concentration in the solvent.}
\end{figure}

To begin, we consider the structures formed when various
concentrations of the sphere-forming copolymer
PCL$_\text{5k}$PEO$_\text{2k}$  are added to the long
lamella-former PCL$_\text{10k}$PEO$_\text{2k}$. These molecules are
chosen to isolate the effect of molecular weight on the shape
transitions: they contain more monomers than the
PCL$_\text{5k}$PEO$_\text{1k}$ copolymers studied previously, but have
the same ratio of hydrophobic to hydrophilic blocks. We first turn our
attention to the turbidity traces of this system. The features in a turbidity trace can be directly linked to the points
where the transitions between spherical micelles, wormlike micelles
and vesicles occur \cite{adams}. Specifically, in clear solutions
spherical micelles or short worms dominate and no or very
few vesicles are present. Conversely, high turbidity of the solution
at high water content indicates the presence of larger aggregates such
as vesicles \cite{adams}. From these
measurements (Figure \ref{turbidity1_fig}) it can be seen that a mixture of
$10\%$ PCL$_\text{5k}$PEO$_\text{2k}$ with $90\%$
PCL$_\text{10k}$PEO$_\text{2k}$ does not behave very differently from
a system of pure lamella-former, since the traces
for the two systems are very similar. However, a sharp change in the optical
transmission is seen on addition of $20\%$
PCL$_\text{5k}$PEO$_\text{2k}$. Here, although the concentration of
sphere-former is still relatively low, the trace resembles that of pure
PCL$_\text{5k}$PEO$_\text{2k}$ much more closely than that of the pure
lamella-former PCL$_\text{10k}$PEO$_\text{2k}$ and does not show the strong turbidity
in the water-rich area linked with the presence of larger aggregates \cite{adams}. We note in passing that the sharp dip in
the optical transmission between $20$ and $30\%$ water is most
probably due to a miscibility gap in the PEO-THF-water phase space
\cite{adams,cristobal,schuhmacher} and is not associated with a
transition in the shape of the aggregates.

These results are in contrast to those obtained for mixtures of
PCL$_\text{5k}$PEO$_\text{1k}$ and PCL$_\text{5k}$PEO$_\text{2k}$ in
our previous publication \cite{schuetz}. There, vesicles were formed up to approximately $30\%$
sphere-former, whereas in the current system the transition from
vesicles to
micelles occurs between $10\%$ and $20\%$
PCL$_\text{5k}$PEO$_\text{2k}$. Increasing the length of the
lamella-forming copolymer is therefore seen to favor the formation of
more curved structures.

\begin{figure}
\includegraphics[width=\linewidth]{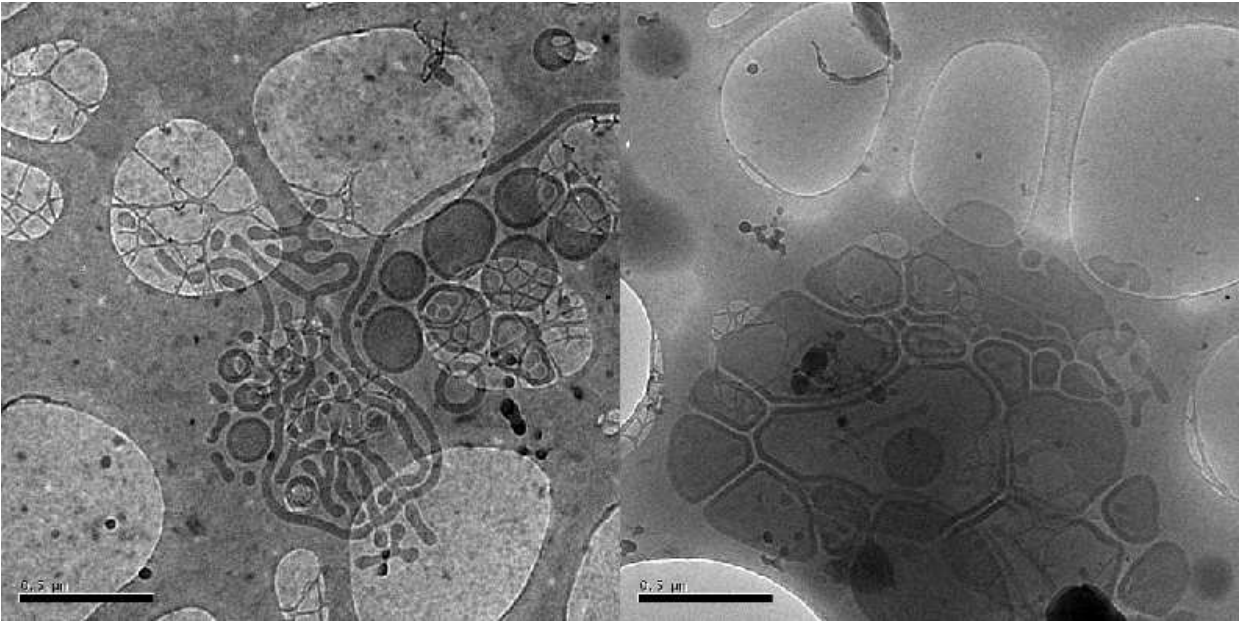}
\caption{\label{TEM1_fig} Cryo-TEM of a mixture of $90\%$ PCL$_\text{10k}$PEO$_\text{2k}$ and $10\%$ PCL$_\text{5k}$PEO$_\text{2k}$ self-assembled by solvent exchange from THF. The images were taken in an aqueous solution with $28\%$ THF; the scalebars are $500\,\text{nm}$. }
\end{figure}

To gain more detailed insights into the system, we now consider
cryo-TEM images taken at a range of sphere-former concentrations.
In mixes with $90\,\%$ PCL$_\text{10k}$PEO$_\text{2k}$ and $10\,\%$
PCL$_\text{5k}$PEO$_\text{2k}$, this technique reveals
the presence of a large variety of structures, as can be seen in Figure
\ref{TEM1_fig}. First, we note that a significant number of vesicles is
still present, accounting for the high turbidity of this mixture seen
in Figure \ref{turbidity1_fig}. In addition, wormlike micelles and rings
(end-to-end joined worms) can be
seen in the left image. The wormlike micelles here often form branched
network structures showing multiple
three-way connections. The individual branches of the network also tend to be rather short, terminating in
enlarged end-caps. These images
closely resemble some of those shown by Jain and Bates
\cite{jain_bates_macro} for PEO-$b$-PB block-copolymer mixtures as
well as those of Chen {\em et al.}
\cite{chen} for PS-$b$-PAA. In different regions of the same TEM grid,
very unusual vesicles could also be seen that on drying deformed to
create a space-filling tessellated pattern (Figure \ref{TEM1_fig}, right
image). The majority of these structures clearly show the dark outer
ring of the vesicle wall confirming that they are indeed closed
vesicles. However, on the top right there are some structures that do
not have this pronounced darker rim and may therefore be unwrapped
bilayer sheets. Such structures have been proposed as intermediate
stages in the self-assembly of vesicles \cite{lasic,antonietti}.

\begin{figure}
\includegraphics[width=\linewidth]{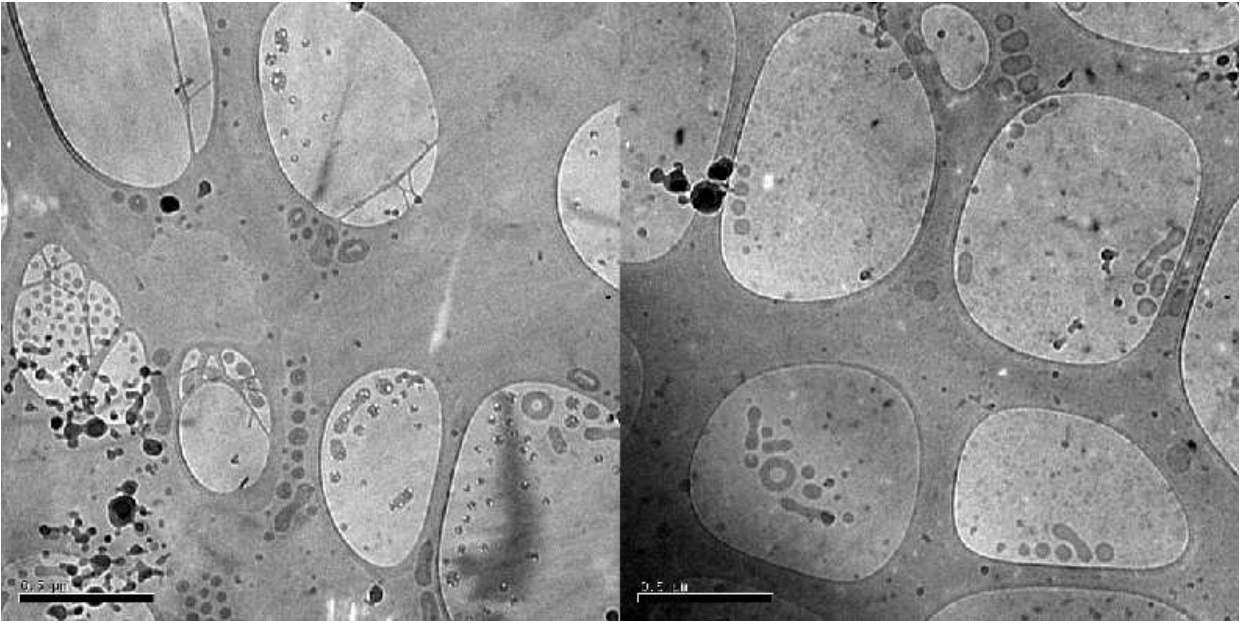}
\caption{\label{TEM2_fig} Cryo-TEM of a mixture of $80\%$ PCL$_\text{10k}$PEO$_\text{2k}$ and $20\%$ PCL$_\text{5k}$PEO$_\text{2k}$ self-assembled by solvent exchange from THF. The images were taken in an aqueous solution with $28\%$ THF; the scalebars are $500\,\text{nm}$.}
\end{figure}

At a mixing ratio of $80\%$ PCL$_\text{10k}$PEO$_\text{2k}$ and $20\%$
PCL$_\text{5k}$PEO$_\text{2k}$ (Figure \ref{TEM2_fig}), the TEM shows a mix of
spherical micelles, short worms and toroidal rings. The vesicles and
sheet-like structures shown in Figure \ref{TEM1_fig} no longer appear at this
concentration. This is in line with the turbidity trace results: this
solution is clear at high water concentrations, consistent with the
presence of small micelles such as rings and short rods \cite{adams}.

We now present our self-consistent field theory calculations on our simple
model of our experimental systems. For each set of calculations, we first consider the
PCL$_\text{5k}$PEO$_\text{1k}$-PCL$_\text{5k}$PEO$_\text{2k}$
mixture studied in our previous paper \cite{schuetz}, modeled, as
described in the methods section, by a
polymer blend including relatively short symmetric lamella-formers containing $N/2$
monomers and sphere-formers containing $N$ monomers. Next, we move on
to our model of the
PCL$_\text{10k}$PEO$_\text{2k}$-PCL$_\text{5k}$PEO$_\text{2k}$ system
of the current paper, in which the SCFT lamella-formers are still
symmetric but now contain $3N/4$ monomers, and look at how the self-assembly is
altered by the change in polymer architecture.

To begin, we calculate the free-energy densities of ideal spheres,
infinite cylinders and infinite bilayers using the method of variable subsystem size described above, and determine how these vary as the volume fraction of
sphere-former is increased. To ensure that the system is relatively
dilute and that aggregates are
surrounded by a large volume of solvent, we fix the overall volume fraction of
copolymer to $8\,\%$. All free energy densities $F/V$ are
plotted with respect to that of the homogeneous solution with the same
composition $F_\text{h}/V$; that is, we plot the quantity
$f=F/V-F_\text{h}/V$. Since the free-energy densities $f_i$ of the
three shapes of aggregate are quite close together, they are plotted
normalized with respect to the magnitude of the free-energy density
$|f_\text{C}|$ of the cylindrical micelle to show the shape
transitions clearly. The cylinder
free-energy density then appears as a horizontal line at $f=-1$, and
is approached from above and below by the sphere and lamella
free-energy densities as the sphere-former concentration increases.

\begin{figure}
\includegraphics[width=\linewidth]{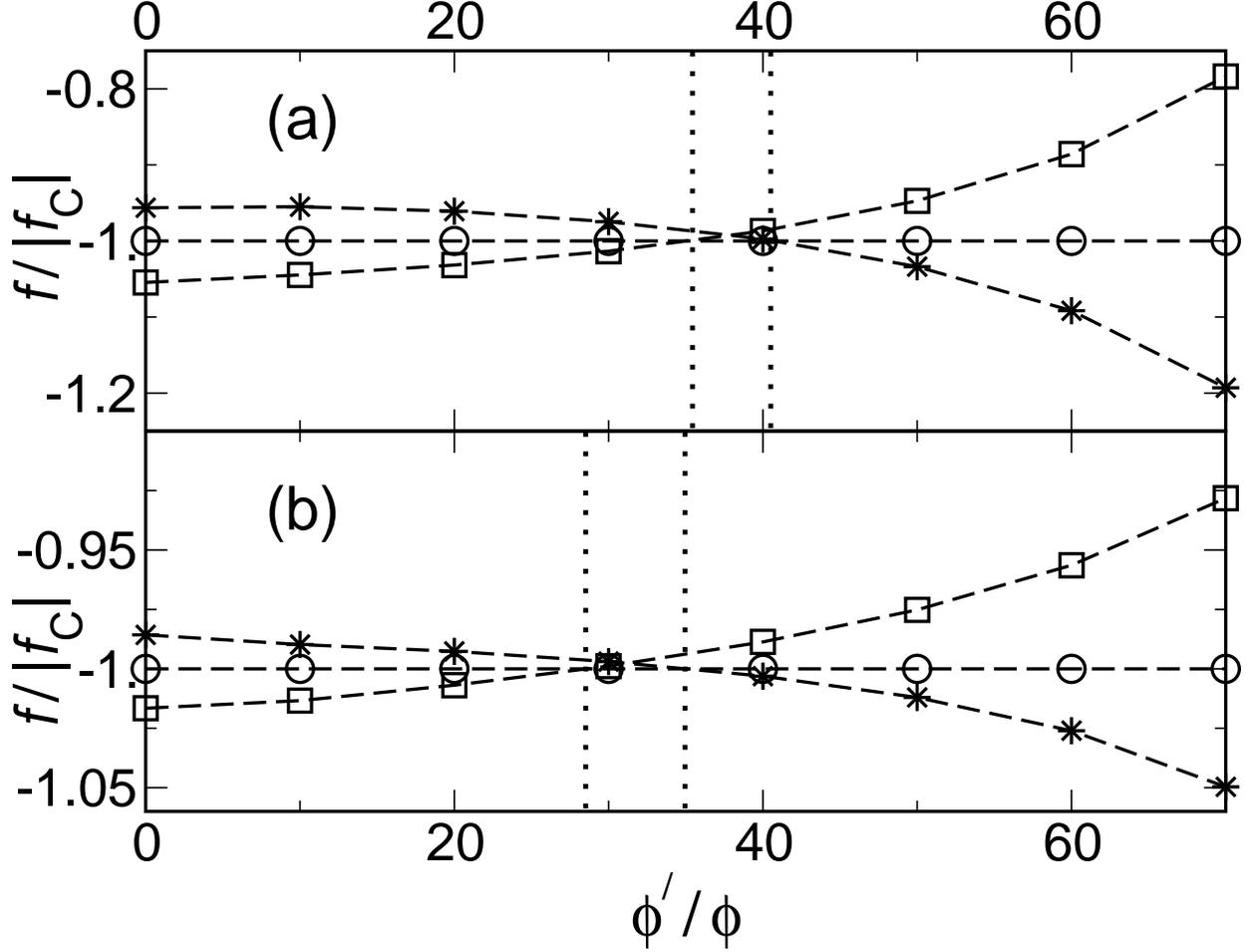}
\caption{\label{phase_fig} Free-energy densities as a function of
  mixing ratio for (a) a short symmetric lamella-former with $N/2$ monomers
  blended with a $N$-monomer sphere-former  and (b) a longer symmetric lamella-former with $3N/4$ monomers
  blended with a $N$-monomer sphere-former. The free-energy densities
  for spherical micelles, cylindrical micelles and flat bilayers are
  plotted with stars, circles and squares respectively. The
  transitions between the different micelle morphologies are indicated
by vertical dashed lines.}
\end{figure}

Figure \ref{phase_fig} shows the free-energy densities of spherical, cylindrical
and planar aggregates plotted against $\phi'/\phi$, where $\phi'$ is
the volume fraction of sphere-former and $\phi$ is the
total volume fraction of copolymer (sphere formers plus
lamella-formers). Panel a shows the results for the system with the
shorter
lamella-former of $N/2$ monomers \cite{schuetz}. In this system, the
lamella-former has the lowest free energy at lower sphere-former
volume fractions. At around $\phi'/\phi=35\%$,
the lamellar and cylindrical free energies cross, and the cylinder has the lowest free
energy until $\phi'/\phi\approx 40\%$, when the spherical micelle finally
becomes most energetically favorable. This reproduces the
series of transitions from vesicles to cylindrical micelles (worms and rings)
to spherical micelles seen in our experiments \cite{schuetz}, although
the values of $\phi'/\phi$ at which the transitions occur are slightly
shifted, as would be expected in view of the simplicity of our model. In these TEM images
\cite{schuetz},
vesicles were seen at $5\%$ and $25\%$ sphere-forming copolymer and spherical
micelles are seen at $75\%$ sphere-former, in line with our
calculations. However, a mixture of worms, rings and vesicles is seen at
$50\%$, where our calculations predict spherical micelles. It is
interesting to note that both our calculations and experiments demonstrate
that a blend of sphere-forming and lamella-forming amphiphiles can
form cylindrical micelles, even though this structure is favored by
neither of these molecules individually \cite{gg}.

In panel b of Figure \ref{phase_fig}, we plot the corresponding free-energy
densities for the same system but with the degree of polymerization of the symmetric
lamella former increased to $3N/4$ monomers: a simple representation of the longer
PCL$_\text{10k}$PEO$_\text{2k}$ molecules of the current
system. Again, we find the same sequence of morphologies as the
sphere-former concentration is increased. However, as in the
experiments, the vesicle-cylinder and cylinder-sphere transitions are
shifted to lower sphere-former volume fractions with respect to the
previous system \cite{schuetz}. Specifically, the
former transition now occurs at approximately $\phi'/\phi=28\%$, whilst
the latter is moved to around $\phi'/\phi=35\%$. Again, due to the
simplicity of our theoretical approach, these transitions are not
perfectly aligned with those in the experimental system. For example,
the TEM images taken at a mixing ratio of $20\%$ (Figure \ref{TEM2_fig}) show a
mixture of rings, short rods and spherical micelles, whereas our
theory predicts that this mixture will form vesicles, or,
equivalently, that the region of coexistence of morphologies would be
expected at a mixing ratio closer to $30\%$.

We believe that the shift in the shape transitions to lower sphere-
former volume fractions for increasing chain length of the
lamella-former is primarily due to the greater shape asymmetry of
the lamella former when its length is increased whilst keeping the
ratio of the hydrophilic and hydrophobic blocks constant. This can be
seen from the following heuristic model of the shape asymmetry of
diblock copolymers. We
assume that the hydrophilic block A of the lamella-former is swollen
by solvent
and so has an end-to-end distance \cite{jones_book} of $R_\text{A}\sim
N_\text{A}^{3/5}$ and effective volume $V_\text{A}\sim
R_\text{A}^3\sim N_\text{A}^{9/5}$. In contrast, the hydrophobic
B-blocks are taken to be in a collapsed, brush-like state
\cite{safran_book} and so have
effective volume $V_\text{B}\sim N_\text{B}$. Since we
keep the ratio of the hydrophilic and hydrophobic blocks constant,
we can also write $V_\text{A}\sim N_\text{tot}^{9/5}$ and $V_\text{B}\sim
N_\text{tot}$, where
$N_\text{tot}=N_\text{A}+N_\text{B}$ is the total length of the lamella-former. Defining
the shape asymmetry of the lamella-former to be $\epsilon=V_\text{A}/V_\text{B}$, we
have $\epsilon\sim N_\text{tot}^{4/5}$. Therefore, as we increase $N_\text{tot}$
at constant hydrophilic to hydrophobic ratio, $\epsilon$ will also
increase and the lamella-former will
become more asymmetric. In consequence, we expect it to
have an greater tendency to form curved structures, with the
swollen A-blocks on the outside of the curved surface. This shift
towards more curved aggregates as the overall copolymer length is
increased has indeed been seen in experiments on symmetric diblock
copolymers in solution \cite{kaya}.

\begin{figure}
\includegraphics[width=\linewidth,angle=270]{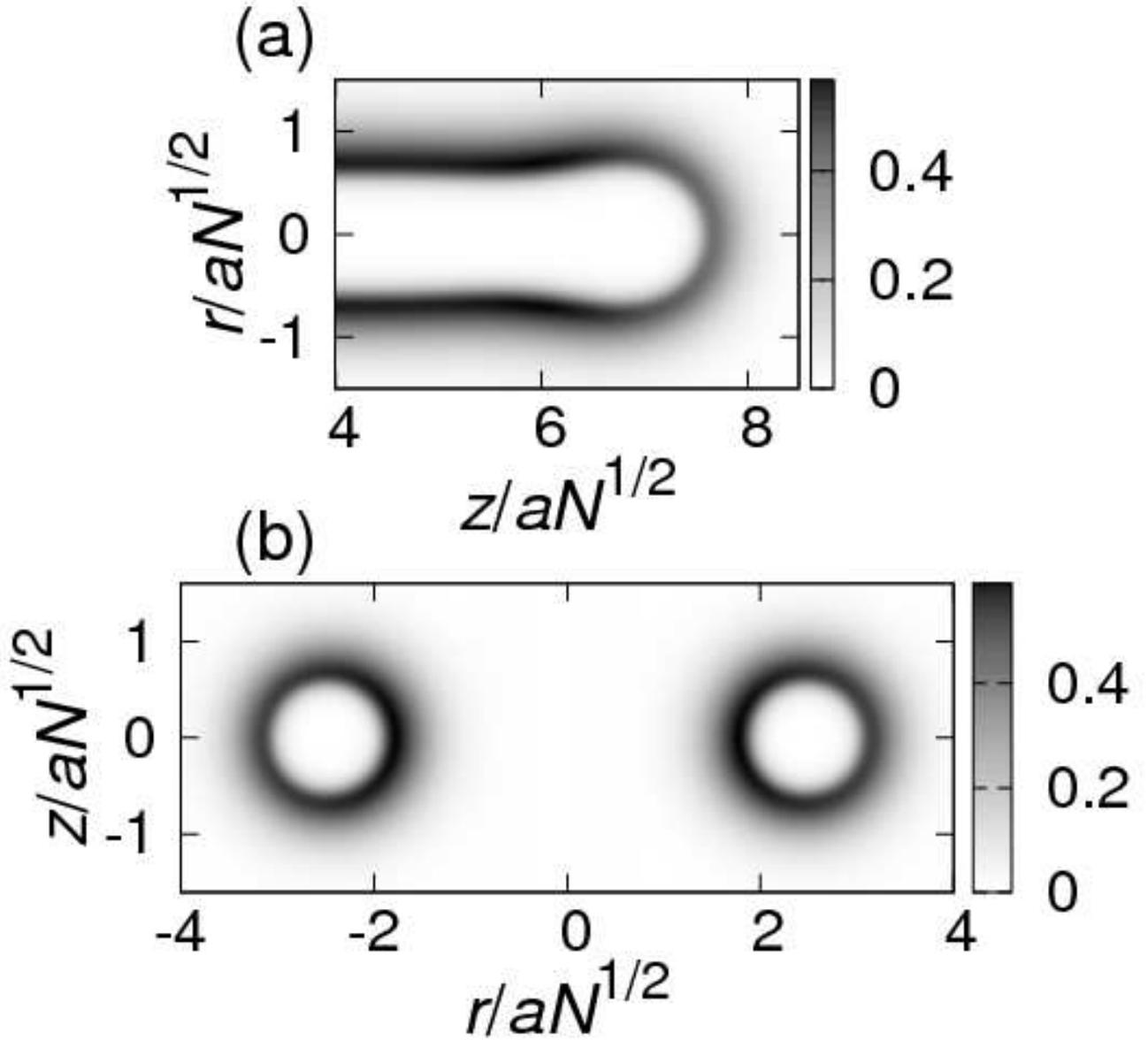}
\caption{\label{phi_A_fig} Bulbous-ended rod (a) and ring (b) structures in a
  system with a long symmetric lamella-former with $3N/4$ monomers
  blended with a $N$-monomer sphere-former in a ratio of 2 to 1. The total
  density of the hydrophilic A-blocks is plotted in cylindrical polar coordinates }
\end{figure}

In the above calculations, we have studied only infinite
cylinders. However, the cylindrical micelles seen in the experiments
have the form of rings or short, bulbous-ended rods. We now use
effectively 2d SCFT calculations in cylindrical polar coordinates
\cite{schuetz,gg} to
study these structures in more detail. First, in Figure \ref{phi_A_fig}, we
demonstrate that these structures can indeed be reproduced within our
model's parameter space. We focus on a system of lamella former with $3N/4$
monomers mixed with sphere-former in a ratio of 2 to 1: a blending
ratio where cylindrical micelles have the lowest free energy in our
calculations (Figure \ref{phase_fig}). To show the form of the structures,
the sum of the hydrophilic block densities of the two species is
plotted. Panel a of Figure \ref{phi_A_fig} shows the bulbous end of a rod
\cite{bernheim-groswasser}, preceded by a thinner neck region of negative
curvature \cite{jodar-reyes_linear,jodar-reyes}. We find also that
ring-like structures exist as local solutions to SCFT, and plot a
cross-section through one of these aggregates in panel b.

\begin{figure}
\includegraphics[width=\linewidth,angle=270]{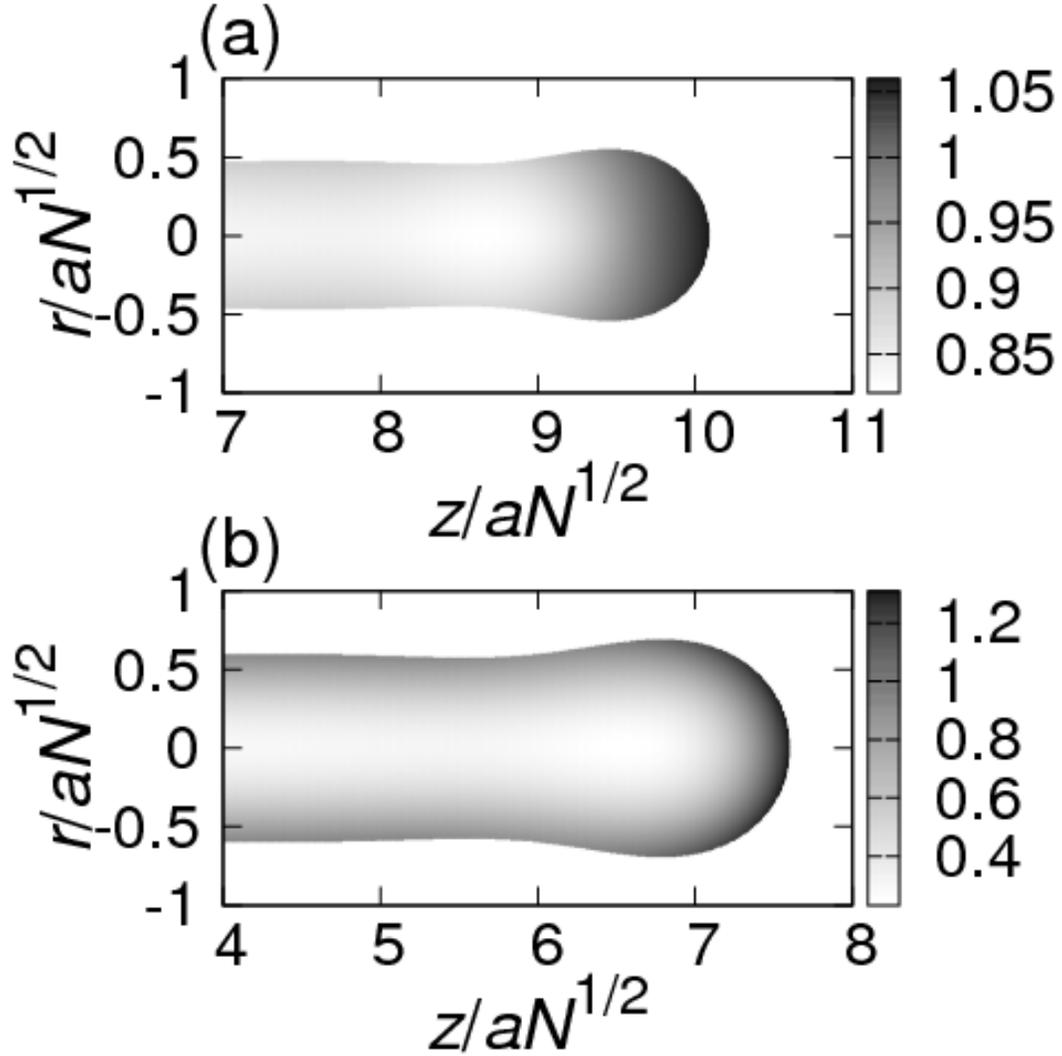}
\caption{\label{eta_fig} Enhancement factor $\eta(\mathbf{r})$ plotted
  within the cores of micelles formed from (a) a 2:1 blend of
  $3N/4$-monomer lamella-former and $N$-monomer sphere-former and (b)
  a 2:1 blend of $N/2$-monomer lamella-former and $N$-monomer sphere-former. Dark
  areas show regions where the concentration of the sphere-former is
  enhanced with respect to the lamella-former, lighter areas show regions where it is depleted. }
\end{figure}

We now turn our attention to the distribution of the two copolymer species
within the rod structures. To this end, we introduce an {\em enhancement
  factor} \cite{schuetz,gg} $\eta(\mathbf{r})$, which we define as
\begin{equation}
\eta(\mathbf{r})=\frac{\phi_\text{B2}(\mathbf{r})}{\phi_\text{B}(\mathbf{r})}\frac{\overline{\phi}_\text{B}}{\overline{\phi}_\text{B2}}
\label{enhance}
\end{equation}
Here, $\phi_\text{B}(\mathbf{r})$ is the local volume
fraction of lamella-former hydrophobic blocks and
$\phi_\text{B2}(\mathbf{r})$ is the corresponding quantity for the
sphere-former hydrophobic blocks. The mean volume fractions of the two
species are denoted by $\overline{\phi}_\text{B}$ and
$\overline{\phi}_\text{B2}$.
The enhancement factor tells us how much the concentration of sphere-former is
enhanced with respect to that of the lamella-former at a given point
in the system. We define $\eta(\mathbf{r})$ such that it is normalized
with respect to the mean
volume fractions of the two core species, so that values greater than one represent
enhancement of the sphere-former concentration and values less than
one represent depletion. The enhancement factor is plotted only
within the core of the micelle, defined as the region where the total
density of the hydrophobic species
$\phi_\text{B}(\mathbf{r})+\phi_\text{B2}(\mathbf{r})$ is greater than
that of the hydrophilic species
$\phi_\text{A}(\mathbf{r})+\phi_\text{A2}(\mathbf{r})$. To locate this
region accurately, we apply simple bilinear interpolation to our SCFT
data to make the grid finer before plotting $\eta(\mathbf{r})$. In
Figure \ref{eta_fig} (a), we show the
enhancement factor for the blend of sphere-former and shorter
lamella-former ($N/2$ monomers) \cite{schuetz}. In this system, the
sphere-formers segregate to the end of the rod \cite{schuetz}, which is the
most strongly curved part of the aggregate and, in fact, closely
resembles a spherical micelle. Since the hydrophobic blocks of the two
species are matched and so can both reach the center of the micelle,
the ratio of the concentrations of the two species varies rather little in the
main cylindrical body of the aggregate away from the endcaps. This is in contrast to the
system shown in panel b of Figure \ref{eta_fig}, where a longer
lamella-former of $3N/4$ monomers is used. Here, the sphere-formers
have shorter hydrophobic blocks than the lamella-formers, and no
longer reach the central axis of the cylindrical micelle. This means
that the sphere-former concentration is enhanced on the surface of the
body of the cylinder as well as in the endcap, and is strongly
depleted in the center of the rod. We note also that this
mismatch between the hydrophobic blocks means that
segregation is a stronger effect than in the previous system, with the
range of values taken by $\eta(\mathbf{r})$ significantly
increased.

\begin{figure}
\includegraphics[width=\linewidth]{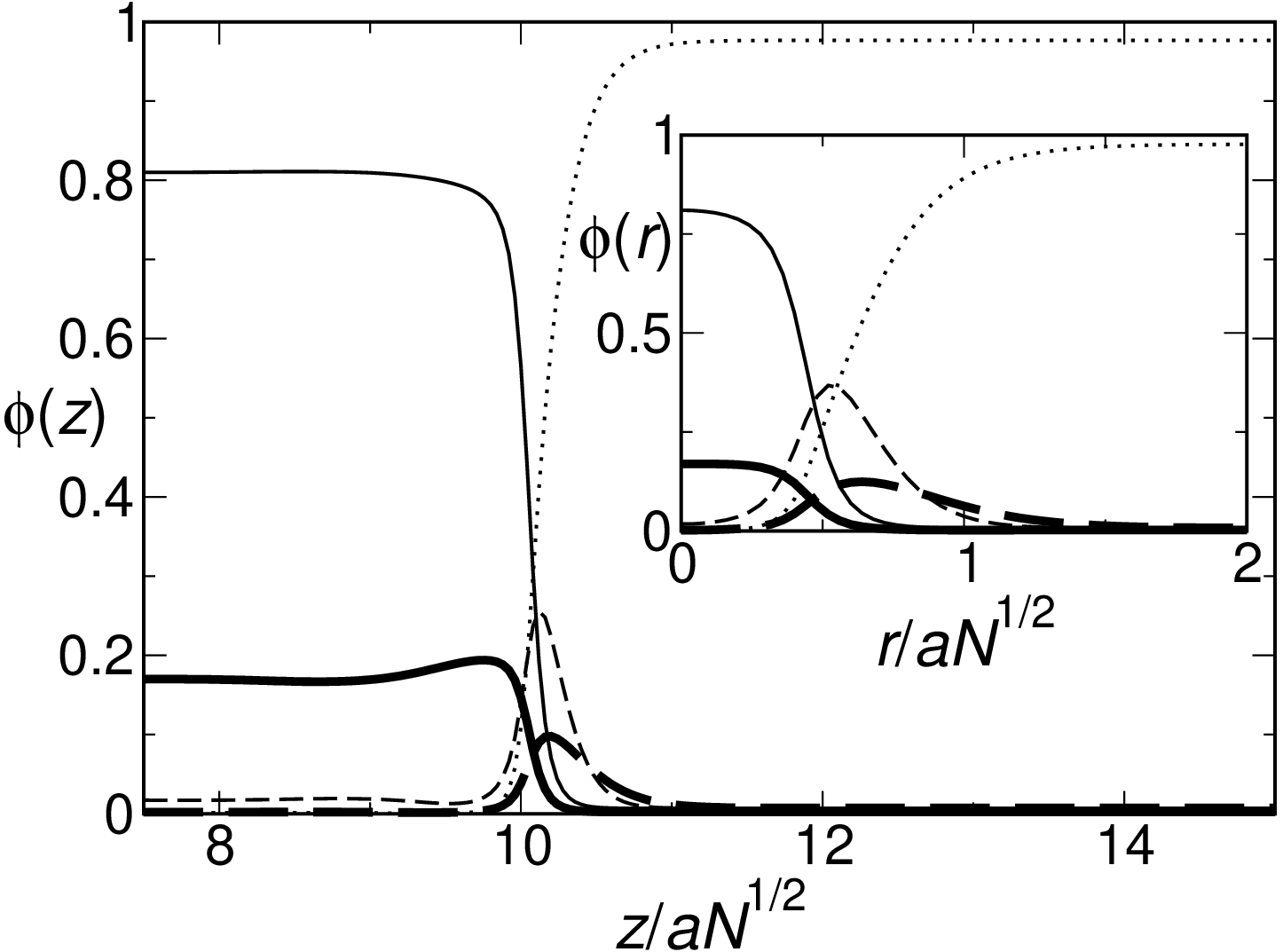}
\caption{\label{cuts1_fig} Cuts through the density profiles of the rod
structure in a 2:1 blend of
  $N/2$-monomer lamella-former and $N$-monomer sphere-former. The
  hydrophobic and hydrophilic block density profiles are shown with
  full and dashed lines respectively, and the sphere-formers are
  plotted with thicker lines. The solvent is plotted with a dotted
  line. The main panel shows a cut along the
  main axis of the rod and the inset shows a cut perpendicular to this
axis at the center of the rod.}
\end{figure}

\begin{figure}
\includegraphics[width=\linewidth]{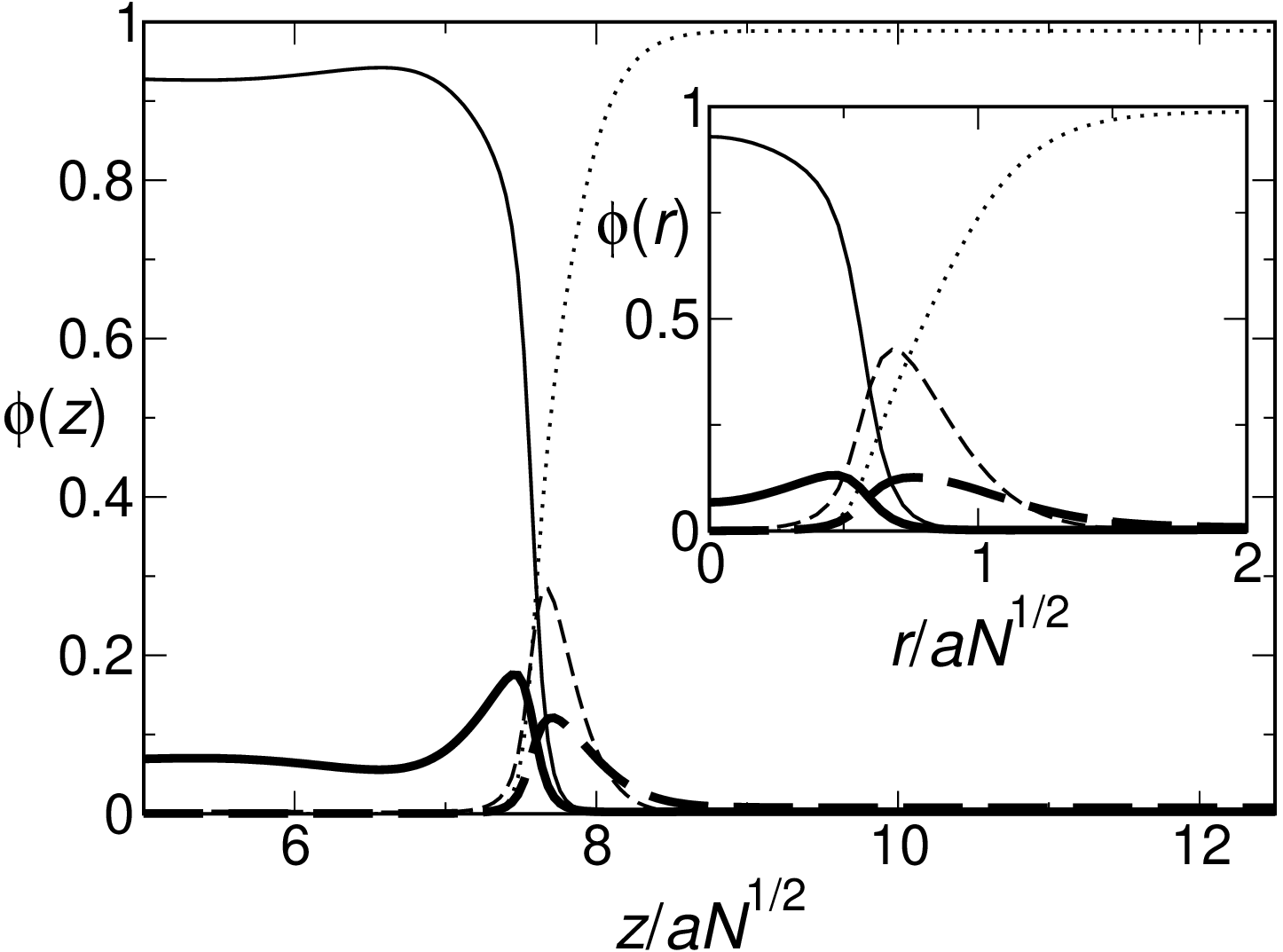}
\caption{\label{cuts_fig} Cuts through the density profiles of the rod
structure in a 2:1 blend of
  $3N/4$-monomer lamella-former and $N$-monomer sphere-former. The
  hydrophobic and hydrophilic block density profiles are shown with
  full and dashed lines respectively, and the sphere-formers are
  plotted with thicker lines. The solvent is plotted with a dotted
  line. The main panel shows a cut along the
  main axis of the rod and the inset shows a cut perpendicular to this
axis at the center of the rod.}
\end{figure}

To study the distribution of the sphere- and lamella-formers within
the micelles in more detail, we now plot cuts
through the density profiles of the various species both along and
perpendicular to the axis of the cylinder for the two systems shown in Figure
\ref{eta_fig}. The plot for the
system with shorter lamella-formers (Figure \ref{cuts1_fig}) confirms two
points made above. First, it can be seen from the cut along the rod axis shown
in the main panel of
Figure \ref{cuts1_fig} that the segregation of the sphere-formers to the endcap
is relatively weak and that the peak in sphere-former concentration here is
quite small. Furthermore, from the inset to Figure \ref{cuts1_fig}, which shows a cut
perpendicular to the rod axis at the center of the micelle, we
see that there is no segregation of the two polymer species in the main body of the
aggregate. 
In corresponding plots for the system with longer lamella-formers
(Figure \ref{cuts_fig}), we see strong segregation of the sphere-formers to
both the endcaps and surface of the cylinder. In addition, plotting
the data in this way also shows us that, despite the new
effect of the enhancement of sphere-former concentration on the
surface of the cylinder, the strongest segregation is still to the
most highly-curved region: the endcaps. To see this, note that the
peak in sphere-former concentration at the end of the rod
(main panel of Figure \ref{cuts_fig}) is higher and more
pronounced than that at the surface of the cylindrical section of the
rod (inset to Figure \ref{cuts_fig}). We note also that the sphere-formers have a dip in
concentration just before the peak at the cylinder endcap. This
feature corresponds to the negatively-curved neck of the micelle \cite{jodar-reyes}
visible in Figure \ref{eta_fig}. Since the sphere-formers naturally prefer
positive curvature, they migrate away from this region.

We believe that the stronger segregation of the sphere-formers in the
mixture with the longer lamella-former is due to the entropic
elasticity of the core blocks. Specifically, if the sphere-formers
were homogeneously mixed with the longer lamella-formers, the
hydrophobic cores would
have to be strongly stretched, restricting the number of
configurations they can access and so leading to a loss of
entropy. In consequence, the sphere-formers move to the endcaps,
leading to the formation of short cylinders or network structures with
many bulbous endcaps such as those seen in Figure \ref{TEM1_fig}.

The larger number of Y-junctions \cite{dan_safran} in the
PCL$_\text{10k}$PEO$_\text{2k}$ system compared to the previous
PCL$_\text{5k}$PEO$_\text{1k}$ system \cite{schuetz} may be explained in a
similar way \cite{jain_bates}: in the system
containing the shorter lamella-former PCL$_\text{5k}$PEO$_\text{1k}$, the relatively short core blocks
can adopt a smaller number of conformations and so are less able to
pack into a more complex structure such as a Y-junction. The increased
number of branched structures at higher molecular weights is in agreement with
the work of several other groups \cite{jain_bates,won,chen,dan} on single-component systems.

\subsection{Blends of two lamella-formers}

Having studied the effect of blending two copolymers that individually
form different structures, we now investigate a system where the two
species are lamella-formers but of different lengths. The two
copolymers considered are the shorter PCL$_\text{5k}$PEO$_\text{1k}$ \cite{schuetz} and the
longer PCL$_\text{10k}$PEO$_\text{2k}$ molecule introduced above. As
in our previous publication \cite{schuetz}, we mixed the two polymers
both before and after their individual self-assembly. 
\begin{figure}
\includegraphics[width=\linewidth]{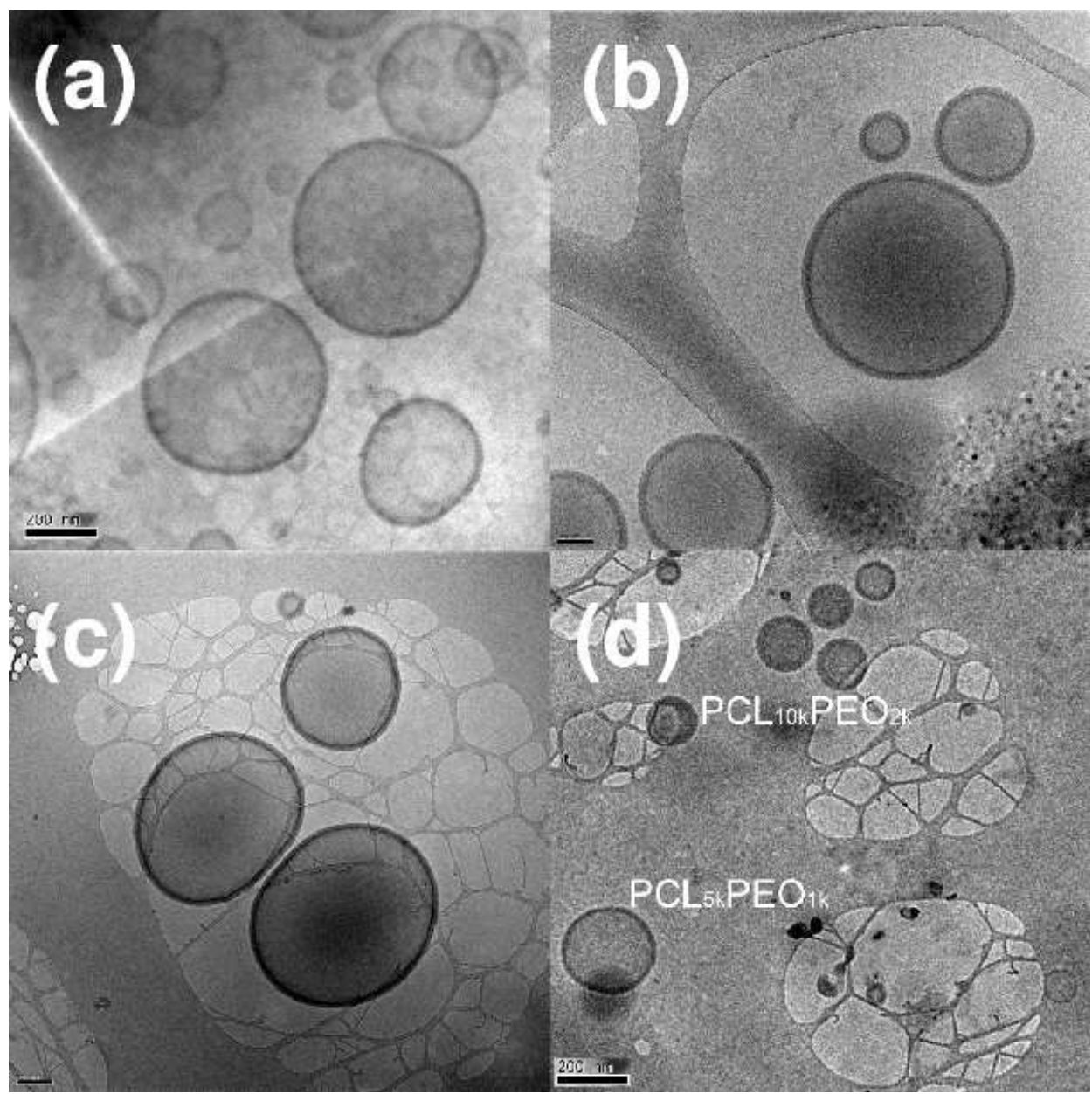}
\caption{\label{TEM3_fig} Cryo-TEM images of the vesicles in $28\,\%$
  THF in water. Panels a and b show the vesicles obtained from the
  self-assembly of PCL$_\text{5k}$PEO$_\text{1k}$ and
  PCL$_\text{10k}$PEO$_\text{2k}$, respectively. Panel c shows the
  structures obtained when solutions of PCL$_\text{5k}$PEO$_\text{1k}$
  and PCL$_\text{10k}$PEO$_\text{2k}$ in pure THF are mixed in equal
  parts prior to self-assembly. Panel d shows the same composition
  when the two solutions are mixed after self-assembly at a THF
  content of $28\%$. The image was taken 1 week after mixing. Two
  distinct populations of vesicles with different wall thickness can
  be seen. All images are at the same magnification. The scalebars in
  a and d are $200\,\text{nm}$ and $100\,\text{nm}$ in panels b and 
c.}
\end{figure}
\begin{figure}
\includegraphics[width=0.6\linewidth]{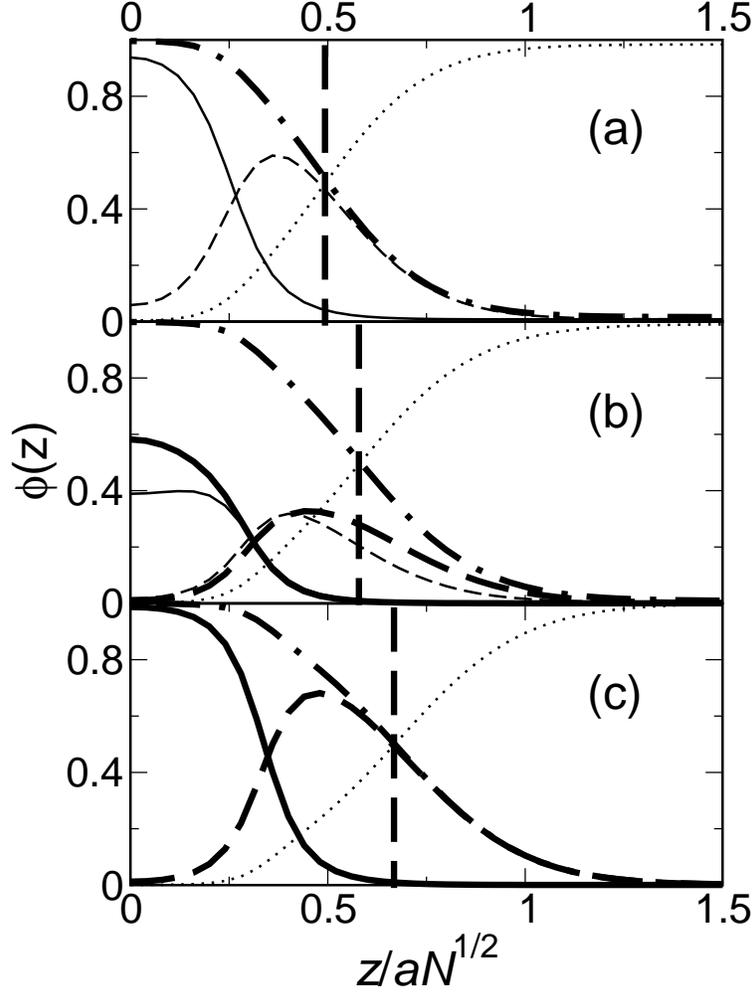}
\caption{\label{lamlam_fig} Density profiles of an infinite lamellar
structure in (a) an $N/2$-monomer lamella-former, (b) a 1:1 blend of
  $3N/4$-monomer lamella-former and $N/2$-monomer lamella-former and
  (c) a $3N/4$-monomer lamella-former.
The
  hydrophobic and hydrophilic block density profiles are shown with
  full and dashed lines respectively, and the $3N/4$ lamella-formers are
  plotted with thicker lines. The solvent is plotted with a dotted
  line, the total density of copolymers is plotted with a
  dot-dashed line, and the point at which these two densities are
  equal is marked with a thick vertical dashed line.}
\end{figure}

Figure \ref{TEM3_fig} shows
representative cryo-TEM images of the vesicles formed by
PCL$_\text{5k}$PEO$_\text{1k}$ (panel a) and PCL$_\text{10k}$PEO$_\text{2k}$
(panel b) on their own. In the second row of the figure the vesicles are
shown that result when the two block-copolymers are mixed before
assembly (in pure THF solution) (panel c) and after assembly i.e.\ after
dilution to $28\%$ THF (panel d). This latter case corresponds to mixing the solutions shown in panels a
and b of Figure \ref{TEM3_fig}. As the chain lengths of the two copolymers
differ by a factor of two, the resulting vesicles have different wall
thickness ($16\,\text{nm}$ and
$25\,\text{nm}$ respectively). In the image d of the sample mixed
post-assembly, the wall thicknesses of the individual vesicles remain
unchanged and two different populations can be clearly distinguished,
while in the sample mixed pre-assembly an intermediate wall thickness
of ca.\ $20\,\text{nm}$ is found. This result indicates that, as
before \cite{schuetz}, the exchange of material between the different
structures is suppressed at THF concentrations below $28\%$.

We now wish to see whether the result of the blending of two types of
lamella-former leading to the formation of vesicles of intermediate
wall thickness can be reproduced in our simple SCFT model. In panel a
of Figure \ref{lamlam_fig}, we plot the density profiles of the hydrophobic B-blocks,
hydrophilic A blocks and solvent of a bilayer formed of
relatively short lamella-formers of $N/2$ monomers calculated, as above, from the
method of variable subsystem size. We also show the total density
profile of the copolymers (A blocks plus B blocks) and the bilayer
thickness, defined as the distance from the origin at which the densities of
copolymer and solvent are equal). Panel c shows the corresponding profiles for the longer
lamella former with $3N/4$ monomers. As expected from the respective
lengths of the molecules, this latter polymer forms thicker
bilayers than those shown in panel a. In the central panel b, we
show the density profiles for a bilayer formed of an equal-parts
mixture of the two species. In excellent
agreement with the experiments, this mixed bilayer has a thickness
approximately halfway between those of the pure bilayers. By comparing
the density profiles in panels a and b we see that
the addition of the longer species (in panel b) causes the core blocks of
the shorter species to
stretch outwards from their equilibrium state in a
single-component structure (panel a).

\subsection{Strongly hydrophobic copolymer mixed with micelle former}

Finally, we investigate an extreme case of two strongly mismatched
copolymers. The first of these, PCL$_\text{5k}$PEO$_\text{550}$, is so
hydrophobic that, in isolation, it fails to assemble into vesicles and
precipitates instead. This polymer was mixed with the micelle-former
considered above, PCL$_\text{5k}$PEO$_\text{2k}$. The two species were
mixed before their individual self-assembly in a mass ratio of $60\%$
PCL$_\text{5k}$PEO$_\text{550}$ to $40\%$
PCL$_\text{5k}$PEO$_\text{2k}$. When self-assembly was triggered by
the addition of water, the solution became turbid, indicating the
presence of large aggregates.
\begin{figure}
\includegraphics[width=\linewidth]{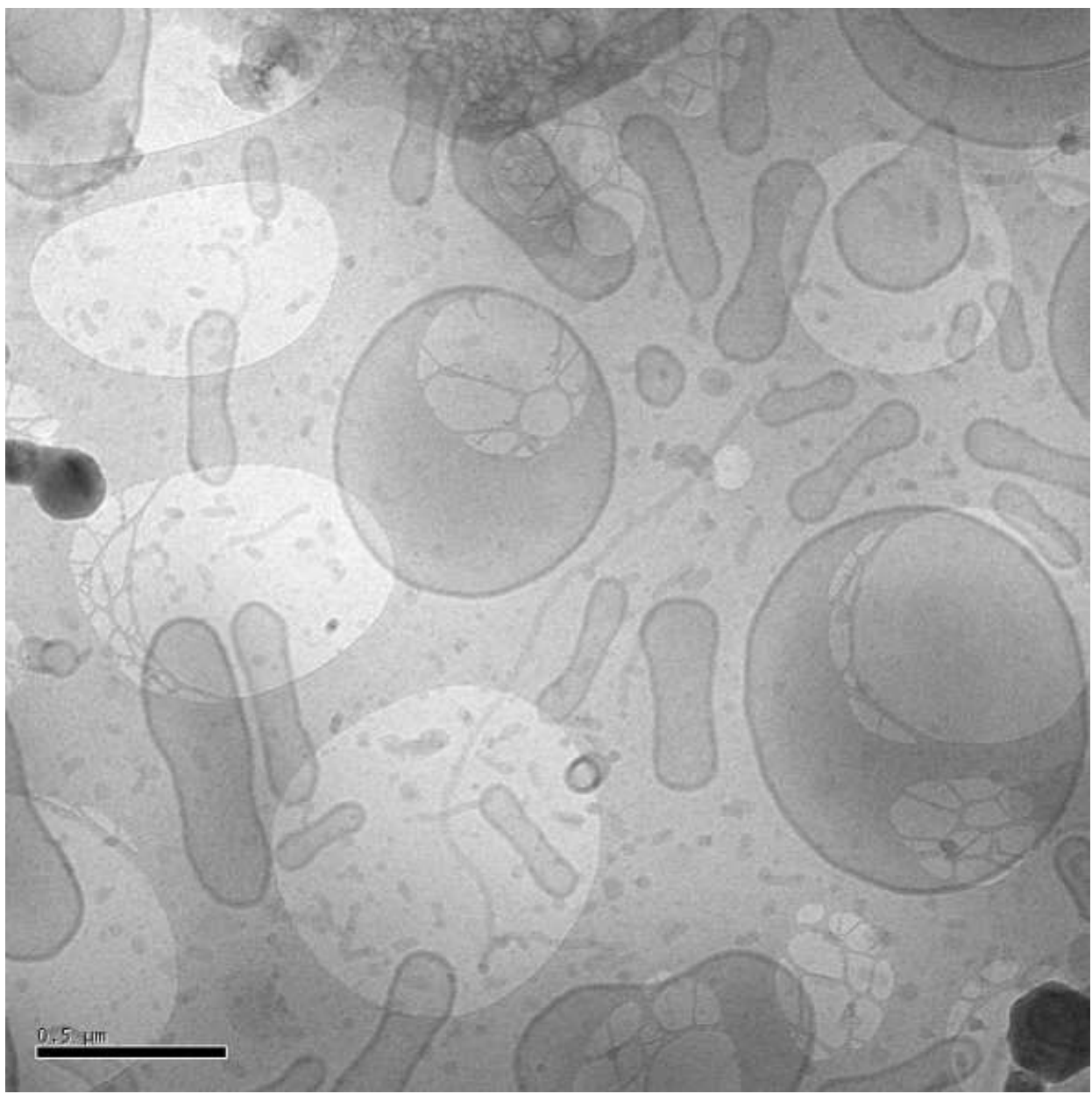}
\caption{\label{blobby_fig} Cryo-TEM of the self-assembled structures obtained for a mixture of PCL$_\text{5k}$PEO$_\text{550}$ and PCL5kPEO2k at a mass-based mixing ratio of 3 to 2 parts. The image was taken from a solution containing $28\%$ THF and the scalebar is $500\,\text{nm}$.}
\end{figure}
This was confirmed by cryo-TEM
(Figure \ref{blobby_fig}), which revealed the presence of a large fraction of vesicles
with smaller populations of wormlike and spherical
micelles. Remarkably, many of these vesicles are not spherical but
have a novel elongated shape, which could be due to
segregation of the two copolymer species within the individual
aggregates. This is likely to be an especially strong effect in the
current system, as the solubilities of the two copolymer species are
very different and the PCL$_\text{5k}$PEO$_\text{550}$ chains start to
aggregate at much lower water fractions than the more hydrophilic
PCL$_\text{5k}$PEO$_\text{2k}$. This produces a phenomenon analogous to
the intermediate mixing discussed in our earlier work \cite{schuetz}:
the strongly hydrophobic chains are partially self-assembled at the
time they encounter the sphere-formers. In consequence, the two
species mix less efficiently, and regions of different curvature can
coexist within the same aggregate \cite{schuetz}. The negative
curvature regions around the centers of the elongated vesicles are likely to
contain higher concentrations of the strongly hydrophobic copolymer,
whereas the curved ends of these structures will probably contain more
of the micelle-former. Due to the small difference in curvature
between the different regions of the aggregate, testing this
hypothesis is unfortunately beyond the scope of our current SCFT methods.

\section{Conclusion}\label{conclusion}
In this paper, we have used a combination of experiment and
theory to show that varying the chain geometry of binary
copolymer mixtures in solution can give us precise control over the
form of the self-assembled aggregates and lead to the formation of new
structures. We investigated three distinct situations. First, we
studied a mixture of sphere-forming and lamella-forming copolymers,
and found in both electron microscopy experiments and coarse-grained
mean-field theory that increasing the chain
length of the lamella-former whilst keeping the ratio of its
hydrophilic and hydrophobic components constant leads to the formation of
highly-curved structures at lower sphere-former volume
fractions. We presented an explanation of this behavior in terms of
the volume asymmetry of the two sections of the diblock. Using more
detailed SCFT calculations, we found
the rings and bulbous-ended rods seen in the experiments, and 
observed a strong segregation of the sphere-forming copolymers to the curved ends
of the cylindrical micelles. We explained this effect by suggesting that sphere-forming copolymers would pay a
large free energy penalty if they were dispersed evenly through the
aggregates, as their relatively short core blocks would need to be
strongly stretched to fit in with those of the lamella-former. In consequence, the sphere-formers tend to de-mix
from the lamella-formers, leading to the formation of cylinders with
highly-curved endcaps. This
segregation between species may be accentuated by other effects such
as enthalpy of crystallization. We also
studied a mixture of two lamella-forming copolymers of different
molecular weights. In both experiment and theory, we found that the bilayers formed in this system
have a wall thickness that is in between those observed in systems
containing only one of the two types of lamella-former. Finally, a mixture of sphere-forming
copolymer and a strongly hydrophobic copolymer that precipitates in
isolation was shown to form a range of structures, including novel
elongated vesicles.

The results presented above demonstrate the power of a combined experimental and theoretical
approach to the investigation and design of self-assembling block
copolymers. Self-consistent field theory can map
out the broad phase diagram of block copolymer mixtures and suggest
experimental parameter spaces to search for new
morphologies. Furthermore, it yields insights into the
aggregates observed in the experiments, reproducing details of the
structures and the distribution of the different polymer species within these.
The wealth of morphologies observed in our work highlights the
fine balance of forces governing the self-assembly behavior of
block-copolymer systems. Further investigation of the different
factors could open up a new zoo of self-assembled aggregates
with the distribution and magnitude of local curvature differences as
additional design parameters, and several avenues for extension of
this work suggest themselves. In particular, the precise role of the
sphere-former architecture could be investigated, with the aim of
producing structures of a specified curvature. The various components
could also be mixed at different stages in their self-assembly
\cite{schuetz}, to access further new aggregates and to gain insight into
the intermediate steps in micelle and vesicle formation. On the theoretical side, our SCFT calculations could
be extended to investigate the favorability of the more complex
structures, particularly the Y-junctions, in different copolymer mixtures.

\section{Acknowledgements}\label{acknowledgements}
This work was performed in Project 264 of the Micro and Nanotechnology
Scheme part-funded by the UK Technology Strategy Board (formerly DTI).
Unilever is thanked for permission to publish this work. MJG is
currently funded by the EU under a FP7 Marie Curie fellowship.

%\bibliography{architecturerefs}

\end{document}